\documentclass[A4]{aa} 

\usepackage{txfonts}
\usepackage{graphicx}
\usepackage[authoryear]{natbib}

\usepackage{longtable} 

\bibpunct{(}{)}{;}{a}{}{,} 

\newcommand{\Teff}{\ensuremath{T_{\mathrm{eff}}}}
\newcommand{\logg}{\ensuremath{\log g}}
\newcommand{\moh}{\ensuremath{[\mathrm{Fe/H}]}}
\newcommand{\dirms}{\ensuremath{\delta I_\mathrm{rms}/I}}
\newcommand{\nobm}{\ensuremath{\mathrm{N}_\mathrm{obm}}}
\newcommand{\lx}{\ensuremath{l_\mathrm{x}}}
\newcommand{\ly}{\ensuremath{l_\mathrm{y}}}
\newcommand{\lz}{\ensuremath{l_\mathrm{z}}}
\newcommand{\Nx}{\ensuremath{N_\mathrm{x}}}
\newcommand{\Ny}{\ensuremath{N_\mathrm{y}}}
\newcommand{\Nz}{\ensuremath{N_\mathrm{z}}}
\newcommand{\vc}{\ensuremath{v_\mathrm{c}}}
\newcommand{\vi}{\ensuremath{v_i}}
\newcommand{\sigx}{\ensuremath{\sigma_\mathrm{x}}}
\newcommand{\varx}{\ensuremath{\sigma_\mathrm{\!x}^2}}
\newcommand{\sigc}{\ensuremath{\sigma_\mathrm{g}}}
\newcommand{\varc}{\ensuremath{\sigma_\mathrm{\!g}^2}}
\newcommand{\sigd}{\ensuremath{\sigma_\mathrm{d}}}
\newcommand{\vard}{\ensuremath{\sigma_\mathrm{\!d}^2}}
\newcommand{\sige}{\ensuremath{\sigma_\mathrm{m}}}
\newcommand{\vare}{\ensuremath{\sigma_\mathrm{\!m}^2}}
\newcommand{\pai}{\ensuremath{P_i}}
\newcommand{\alf}{\ensuremath{\alpha}}
\newcommand{\xmean}[1]{\ensuremath{\left\langle#1\right\rangle}}

\begin{document}

\title{Gravitational redshifts in main-sequence and giant stars  
\thanks{Based on observations collected at ESO, La Silla, Chile, during the agreement between the
Observatorio Nacional at Rio de Janeiro and  ESO.}
} 
  

\author{L. Pasquini\inst{1} \and C.Melo\inst{1} \and C. Chavero \inst{2} \and D. Dravins\inst{3} \and 
H.-G. Ludwig\inst{4,6} \and P.Bonifacio\inst{4,5} \and R. De La Reza \inst{2}  }

\offprints{L. Pasquini, \email{lpasquin@eso.org}}

\institute{ESO -- European Southern Observatory, Karl-Schwarzschild-Strasse 2, 85748 Garching bei M\"unchen, Germany
 \and Observat\'orio Nacional, Rua General Jos\'e Cristino 77, 20921-400 Rio de Janeiro, RJ, Brazil 
 \and Lund Observatory, Box 43, 22100 Lund, Sweden 
 \and GEPI, Observatoire de Paris, CNRS, Universit\'e Paris Diderot, Place Jules Janssen, 92190 Meudon, France
  \and Istituto Nazionale di Astrofisica, Osservatorio Astronomico di Trieste, Via Tiepolo 11, 34143 Trieste, Italy
  \and ZAH -- Landessternwarte, K\"onigstuhl 12, 69117 Heidelberg, Germany
}

\date{Received  / Accepted }

\abstract
{ Precise analyses of stellar radial velocities reveal intrinsic effects  causing wavelength shifts of spectral lines (other than Doppler shifts due to radial motion), such as gravitational redshifts and convective blueshifts. }
{ Gravitational redshifts in solar-type main-sequence stars are expected to be some 500~m~s$^{-1}$ greater than those in giants.  Such a signature is searched for between groups of open-cluster stars which share the same average space motion and thus have the same average Doppler shift. } 
{ 144 main-sequence stars and cool giants were observed in the M67 open cluster using the ESO FEROS spectrograph, obtaining radial velocities by cross correlation with a spectral template.  Binaries and doubtful members were removed, and averaged low-noise spectra calculated for different classes of stars. }
{ M67 dwarf and giant radial-velocity distributions are well represented by Gaussian functions, sharing the same apparent average radial velocity within $\simeq$~100~m~s$^{-1}$.  In addition, dwarfs in M67 appear to be dynamically hotter ($\sigma$ = 0.90 km~s$^{-1}$) than giants ($\sigma$ = 0.68 km~s$^{-1}$). }
{ Explanations for the lack of an expected signal are sought: a likely cause is the differential wavelength shifts produced by different hydrodynamics in dwarf and giant atmospheres.  Radial-velocity differences measured between unblended lines in low-noise averaged spectra vary with line-strenth: stronger lines are more blushifted in dwarfs than in giants, apparently 'compensating' for the gravitational redshift. 

Synthetic high-resolution spectra are computed from 3-dimensional hydrodynamic model atmospheres for both giants and dwarfs, and synthetic wavelength shifts obtained.  In agreement with observations, 3D models predict substantially smaller wavelength-shift differences than expected from gravitational redshift only.  The procedures developed could be used to test 3D models for different classes of stars, but will ultimately require high-fidelity spectra for measurements of wavelength shifts in individual spectral lines. }

   
\titlerunning{Gravitational redshifts in dwarfs and giants}
\authorrunning{L. Pasquini et al.}
\maketitle

\section{Introduction}
\label{sec:Intro}

\subsection{Gravitational redshift}

Spectroscopic measurements of stellar radial velocities have reached levels of precision which not only enable new types of studies, but also require a better understanding of effects other than a star's barycentric motion that may cause displacements of spectral-line wavelengths.  One such effect is the
gravitational redshift originating from the propagation of light between different gravitational potentials at the source and at the observer.

Any theory involving conservation of energy requires that radiation leaving a gravitational field loses energy, and with the early 20$^\mathrm{th}$ century understanding of the photon and the relation between its energy and wavelength, predictions could be made of the wavelength displacement expected for radiation leaving various stars, and these effects have been sought observationally.  

While a determination of an absolute wavelength shift requires the true stellar center-of-mass motion (and its induced Doppler shift) to be known, {\it relative shifts} can be determined between stars sharing a common space motion.  This applies to the orbit-averaged velocity of binary stars (where each component shares the same systemic velocity), and for stars in [open] clusters which share the same space velocity vector.

The first claimed confirmation of the predicted gravitational redshift came from the measurement of the apparent radial velocity of Sirius B, the white-dwarf companion to Sirius (Adams 1925), identified as the difference from the motion expected in its 50-year orbit.  However, this faint white dwarf is overwhelmed by scattered light from the much brighter primary and the validity (and even the honesty!) of these early measurements has been questioned (Hetherington 1980; Greenstein et al.\ 1985).

Gravitational redshifts have since been more reliably determined for numerous white dwarfs in binary systems and in nearby open clusters such as the Hyades and the Pleiades, finding values typically on order 30--40~km~s$^{-1}$, as measured against wavelength positions in the spectra of ordinary stars.  Also, very much larger gravitational redshifts have been identified from spectra of X-ray bursts on neutron-star surfaces.

The gravitational redshift is proportional to the gravitational potential at the stellar surface and scales proportional to the stellar mass, and inversely as its radius $R$: 

$v_{\rm grav} = GM/Rc$, 

where $G$ is the gravitational constant.  The expected solar gravitational redshift for light escaping from the solar photospheric surface to infinity is 636.486 $\pm$ ~0.024~m~s$^{-1}$ (Lindegren \& Dravins 2003) and -- since the shift diminishes with distance from the stellar center as $r^{-1}$ -- becomes 633.5~m~s$^{-1}$ for light intercepted at the Earth's mean distance from the Sun ($r = 215R_\odot$).  A solar spectral line instead formed at chromospheric heights (30~Mm, say; $r = 1.04R_\odot$) will have this redshift decreased by some 20~m~s$^{-1}$, and a coronal line by perhaps 100~m~s$^{-1}$.

For other stars, the shift scales as $(M/M_\odot)(R/R_\odot)^{-1}$, or as $(g/g_\odot)(R/R_\odot)$, where $g$ is the surface gravity.  For distant stars, additional lineshifts may originate from large-scale gravitational fields in the Milky Way Galaxy.  For instance, spectra of stars in the central Galactic bulge ($R_\ast \simeq 1$~kpc) may be gravitationally redshifted by 300--400~m~s$^{-1}$, while stars in the Magellanic Clouds ($R_\ast \simeq 55$~kpc) might be blueshifted by a similar amount, as seen by an observer near the solar galactocentric position at $R_{\rm obs}\simeq 8.5$~kpc (Lindegren \& Dravins 2003).

Across the Hertzsprung-Russell diagram, the gravitational redshifts are not expected to vary much along the main sequence between A5 V and K0 V (being around 650~m~s$^{-1}$), but to reach twice that value for more massive early-B stars with $M\simeq$~10$M_\odot$.  For giants the shifts decrease substantially and for red supergiants it should be only some 20 or 30~m~s$^{-1}$, three orders of magnitude smaller than for white dwarfs (Griffin 1982; Dravins 2005).  Thus, the difference between the shifts in dwarf and giant stars is expected to be some 0.5~km~s$^{-1}$, orders of magnitude greater than current measurement precision.  Such values are significant, both to study the precise dynamics in binary- and multiple-star systems (Griffin 1982, Pourbaix et al. 2002), or to disentangle wavelength shifts caused by stellar atmospheric dynamics by convective motions (e.g., Dravins 1982; Asplund 2005; Dravins et al.\ 2005).  Together with photospheric line asymmetries, such intrinsic wavelength shifts make up significant diagnostic tools for probing the 3-dimensional structure of stellar atmospheres (Dravins 2008; Nagendra et al.\ 2009), a field where additional diagnostics are needed to reliably segregate among hydrodynamic models predicting significantly different chemical compositions, oscillation properties, and so on.

\subsection{Searching for shifts in ordinary stars} 

At first sight, the Sun would appear to be a promising source for determining absolute gravitational redshifts, since the radial Sun-Earth motion is both very small and well known.  However, already long ago, spectral lines in the Sun were found to be shifted by different but comparable amounts, including variation of absorption-line wavelengths across the solar disk from its center towards the limb that was initially a mystery.  This is now well understood as convective lineshifts, originating from a summation of contributions from hot and rising (thus locally blueshifted) elements, and cool and sinking ones, and various effects of observational perspective across the disk.  Many efforts were made to measure the solar gravitational redshift (and several publications claimed its detection) although, in hindsight, almost all such studies were constrained by the limited understanding of the interplay with other mechanisms causing subtle wavelength shifts (besides effects from atmospheric motions, also the limited accuracy for laboratory wavelengths, and even effects of pressure shifts come to play).  Possibly, the most convincing demonstration of a gravitational redshift in the solar spectrum was by Beckers (1977).  The magnetic fields in the cores of sunspots inhibit convective motions and greatly reduce the amount of both convective wavelength shifts, and their center-to-limb variation, enabling rather precise wavelength shift measurements in magnetically insensitive lines.  From such lines, Beckers deduced a shift of 613~m~s$^{-1}$, consistent with the expected magnitude.

Just as for white dwarfs, systematic wavelength displacements between ordinary stars may be identified in binaries sharing a common systemic velocity, and for stars in open clusters, moving in space with sufficiently small velocity dispersion.  The earliest such attempts seem to have been by E. Finlay-Freundlich who, starting around 1914, tried to find statistical evidence of gravitational redshifts (for a lucid review, see Hentschel 1994).  Somewhat later, Trumpler (1935), recognized that massive O stars should be excellent targets, because they have the largest gravitational redshift while they often appeared in young clusters where they, being massive, by equipartition of energy should have the smallest velocity dispersion.  In some clusters, shifts of some 10~km~s$^{-1}$ were found relative to comparison stars, suggesting enormous stellar masses.   A later understanding of these stars showed that shifts should be much less (Conti et al.\ 1977), and while effects of stellar variability, line asymmetries, and differences of apparent velocities in different spectral features did begin to get appreciated, also a certain literature grew on suggested 'anomalous' redshifts in binary high-luminosity stars, and other objects.

Only after measuring precisions for individual stars had reached levels of around 100~m~s$^{-1}$, did it become practical to search for systematic differences in apparent velocities among ordinary stars, e.g., between main-sequence dwarfs and giants.  One hint came from a study of the open cluster NGC~3680 by Nordstr\"{o}m et al.\ (1997), who found the giants (though the sample contained only six such stars) to be blueshifted on average 0.4~km~s$^{-1}$ relative to the dwarfs, as expected from gravity.

For certain groups of stars, including those in undispersed moving clusters, {\it absolute} stellar radial motion can be determined from precise astrometry, not invoking any spectroscopy.  The difference to spectroscopic lineshifts then reveals shifts intrinsic to stellar atmospheres.  In the Hyades (Madsen et al.\ 2002), such signatures were seen for some classes of hotter main-sequence stars, but the spectral displacements of the (only three) Hyades K-giants did not show any significant deviation from that of main-sequence K-type stars.

Different gravitational redshifts in dwarf and giant stars have thus been sought for almost a century.  The value and usefulness of such measurements has been pointed out from time to time (e.g., von Hippel 1996) but conclusive or statistically significant observations are still lacking, which was a motivation for the present project.

\section{Radial velocities in open clusters}

The aim is to observe apparent radial velocities of main-sequence and giant stars in some stellar cluster(s) with sufficient precision to identify expected differences in gravitational redshift between these groups.  The ideal cluster should obviously be populous and have stars covering a large range of $M/R$ ratios.

Not all clusters are suitable.  To assure good measuring precision for spectroscopic radial velocity, the stars must have spectral lines that are both narrow and numerous, which excludes young clusters where main-sequence stars still are rapidly rotating and produce broadened lines.  Also, higher-temperature giants are unfit due to their few spectral lines and often complex atmospheric dynamics.  Thus, the cluster must be sufficiently old, with main-sequence rotation slower than some 5--10~km~s$^{-1}$, and must have a well developed red-giant branch with many cool giants that are good targets for radial-velocity measurements (e.g., Setiawan et al.\ 2004).  

The relatively old cluster M67 (NGC~2682) at $\sim$850 pc distance was chosen as a nearly ideal object.  Besides having a well-developed red-giant branch, advantages include that its composition is quite similar to solar (Tautvai{\v s}ien{\. e} et al.\ 2000; Randich et al.\ 2006; Pace et al.\ 2008) and the age somewhat comparable: 2.6~Gy (WEBDA 2009).  It has been well studied in various aspects (Pasquini et al.\ 2008; Sandquist 2004), including a census of binaries (Latham et al.\ 1992).  Finally, we had already observed a homogeneous group of stars in M67 (Melo et al.\ 2001; hereafter Paper I), verifying the modest rotational velocities, and could use these data as a starting point.

From previous radial-velocity data (e.g., Mathieu et al.\ 1986), the internal velocity dispersion of the space motions within M67 was found to be $\approx$~0.5--0.8~km~s$^{-1}$, similar to that deduced from proper motions (Girard et al.\ 1989).  Other studies suggest the expansion and rotation of the cluster core to be below 1~km~s$^{-1}$ (Zhao et al.\ 1996; Loktin 2005).

Since these numbers are comparable to the expected difference in gravitational redshifts between dwarfs and giants, averages over many stars will be required, and attention must be paid to calibrating radial-velocity measures between stars of different spectral types, identifying stellar multiplicity, possibly different velocities for stars mass-segregated into different parts of the cluster, and the physics of line formation in dynamic stellar atmospheres.  In any case, since the internal measuring precision of modern radial-velocity instruments is very much better than this dispersion, significant progress should be possible.

\section{Observations and data reduction}
\label{sec:Obs}

An observing program was carried out with FEROS ({\it Fiberfed Extended Range Optical Spectrograph}; Kaufer et al.\ 1999) in three different runs.  The first took place during FEROS commissioning in October and November 1998, observing 28 main-sequence, turnoff, and giant stars.  The radial velocities and $V\sin i$ were published in Paper I and are listed in the online Appendix for sake of completeness.  Additionally 68, mostly main-sequence and turnoff stars, were observed over four nights during the first week of February 2002 (Run II) in parallel with other projects during Brazilian time.  Finally in Run III, 66 additional M67 members were observed in late February 2008.  For the first two runs, the spectrograph was attached to the ESO 1.5-m telescope, and during the third to the MPG/ESO 2.2-m telescope. In total 162 observations for 144 different stars were obtained.

FEROS is a bench-mounted, thermally controlled, prism-cross-dispersed \'{e}chelle spectrograph.  It offers rather high resolution (R~$\simeq$~48,000), and in a single recording with 39 \'{e}chelle orders provides almost complete spectral coverage from $\approx$~350--920 nm.  On bright stars, FEROS has been demonstrated to be precise to better than 20~m~s$^{-1}$ (Setiawan et al.\ 2004).  Our targets are substantially fainter, but an exceedingly high precision is not required.  The M67 velocity dispersion is on the order of 1~km~s$^{-1}$, so a precision of a few hundred meters in a single exposure is sufficient.  Exposure times were set to have a S/N around 30 or better.  With this photon noise, the radial velocity (RV) precision limit is of the order 20~m~s$^{-1}$. The total error budget is discussed in Section 6.1.

We aimed to observe many stars with different $M/R$ ratios, to be able to precisely determine mean velocities for different groups of stars.  Known binaries were excluded (M67 is rich in those), although some single-lined spectroscopic binaries likely are still present.  We also observed some stars with deviant velocities which appear to be non-members. The list for the whole sample, including data from Paper I, is given in the Appendix, available online.
\addtocounter{table}{1}


\section{Radial velocities and membership}

Radial velocities were computed as the Gaussian fit to the cross-correlation function (CCF) between the observed stellar spectrum and a spectral template mask.  This binary mask was based on a stellar spectrum, following the recipe in Baranne et al.\ (1979).  As in Paper I, a K0~III synthetic mask (Baranne et al.\ 1996) was used for all stars, assuring the same reference for all spectra.

The mask is composed mostly of metallic lines, whose number is sufficiently high (about 2000 lines in the 500--700 nm wavelength interval used) that we expect that line blends, which vary according to the stellar effective temperature and gravity, are substantially symmetric and do not change the average shift of the spectrum.  The cross-correlation mask is built by selecting the signal from spectral lines below a given residual flux.  This cut is made in order to compromise between the number of points producing the cross-correlation signal and the width of the lines: too high a threshold (including points close to the continuum) might produce too wide CCFs, decreasing the measuring precision (see Baranne et al.\ 1996).  For our mask a value of 0.7 was selected, i.e., only wavelength points with a residual intensity of less than 0.7 in the K0~III reference spectrum contribute to the digital mask.  This implies that the selected lines have relatively large equivalent widths in the M67 giants. 

\begin{figure}
\centering
\includegraphics[width=9cm]{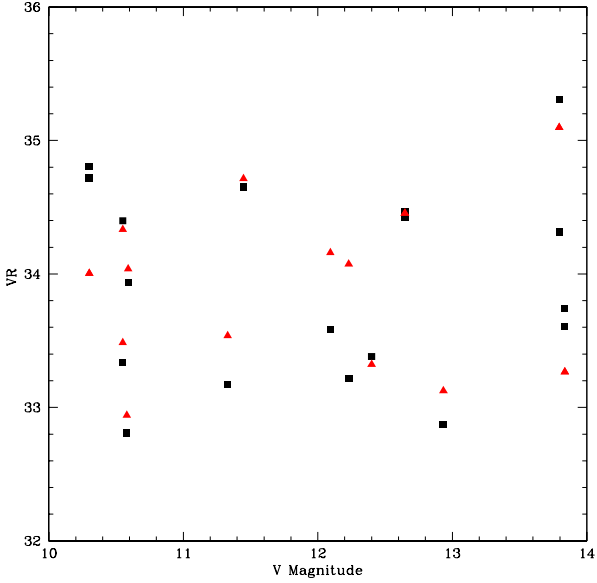}
\caption{ Radial velocities for the 14 stars common to the different observing runs. Observations at the ESO 1.5-m telescope are black squares; those at the MPG/ESO 2.2-m are [red] triangles.  The difference between the averages from the two telescopes is less than 30~m~s$^{-1}$, and the spread is 0.3~km~s$^{-1}$ after 3 suspected binaries are discarded. } 
\label{fig:com_rv}
\end{figure}

Only a few stars have repeated observations, and the comparison between multiple measurements is shown in Figure~\ref{fig:com_rv}.  The average radial velocity of these stars in the first observing runs, made at the ESO 1.5-m, is 33.93~km~s$^{-1}$, which compares very well with 33.96~km~s$^{-1}$ for the same objects observed at MPG/ESO 2.2-m, although the dispersion is quite large, around 0.45~km~s$^{-1}$.  This number is, however, an upper limit to the measurement error, because among the 14 stars observed at the two telescopes, some probably are binaries.  Three stars (S1016, S1075, S1305) differ by more than 0.8~km~s$^{-1}$, and once these are removed from the sample, the dispersion drops to 0.3~km~s$^{-1}$.  We consider this as a good estimate of the real uncertainty of a single measurement, though it might still be an upper limit.  This value is similar to the dispersion in measurements of our RV standard star, HD 10700 (see section 5.1 and 6.1.2).

\begin{figure}
\centering
\includegraphics[width=9cm]{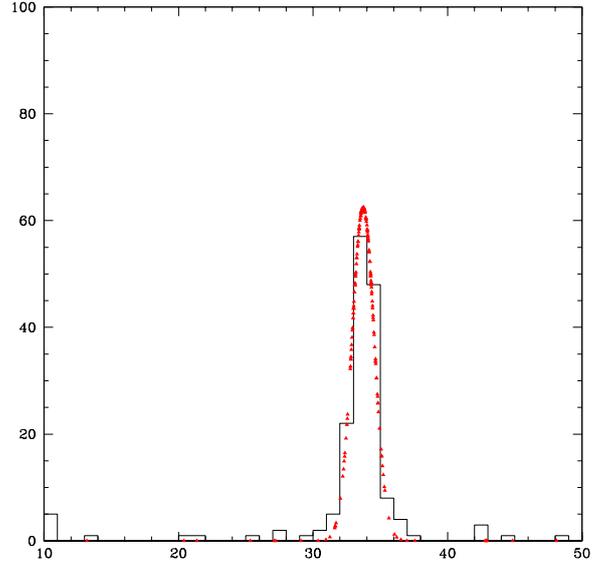}
\caption{Radial velocity (RV) distribution of the observed stars in M67 (continuous line).  A Gaussian distribution centered at Vr=33.73 and with $\sigma$=0.83~km~s$^{-1}$ is superposed}
\label{fig:vrad_distribution_sel}
\end{figure}

\begin{figure}
\centering
\includegraphics[width=9cm]{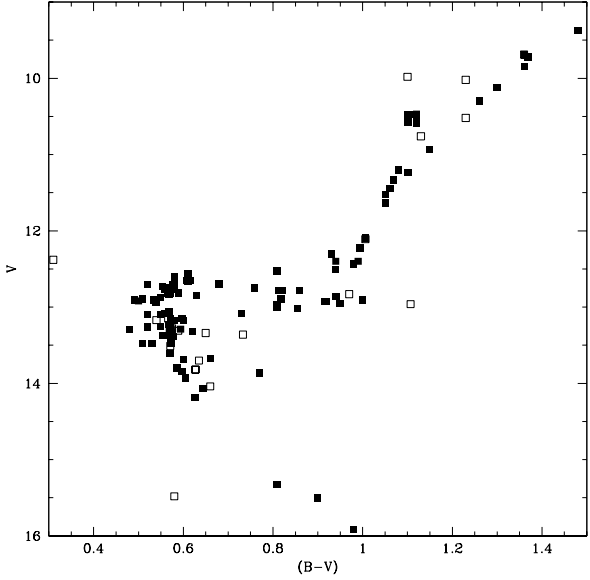}
\caption{M67 color-magnitude diagram with the observed targets. Empty boxes are stars discarded on the basis of their radial velocities differing more than 2.5 $\sigma$ from the average, or known to be binaries. Filled squares denote stars retained as bona fide single members. }
\label{fig:cmd}
\end{figure}

Figure~\ref{fig:vrad_distribution_sel} shows the radial velocity distribution of all the observed stars. The M67 members are clearly visible, and their RV distribution can be well represented by a Gaussian distribution centered on RV=33.73, with $\sigma$=0.83~km~s$^{-1}$.  We made a 2.5 $\sigma$ clipping and considered only the stars left as bona fide M67 single members.  Using this selection, a total of 33 stars were discarded.  We note that S1011, 1064, 1221, 1250, 1314, all classified as binaries in Paper I, appear in the list of binary stars.  Out of the Paper I and Latham et al.\ (1992) lists, only S986 was not recovered as a binary.  The final sample consists of 110 stars and their color-magnitude diagram is in Figure~\ref{fig:cmd}; bona fide single members are marked as filled squares, stars discarded as empty squares.  The targets are well distributed in the color-magnitude diagram, and numeorus evolved stars were observed to ensure good statistics also in that less populated group. 

M67 is a well studied cluster, and among previous radial velocity studies, the most appropriate comparison is probably with the pioneering work by Mathieu et al.\ (1986), which is similar in size to ours (104 likely single members), with comparable (but slightly worse) RV precision.  A difference is that their sample was limited to brighter magnitudes, since it included all proper motion candidates brighter than V=12.8, which corresponds to the region around the turnoff.  Those authors used (as we do) a cross correlation technique, although their spectra were limited to a smaller spectral range, centered on 520 nm. We analyzed the Mathieu et al.\ (1986) radial velocities exactly in the same way as ours, finding a mean velocity 33.57~m~s$^{-1}$($\sigma$= 0.78~km~s$^{-1}$) and, anticipating the results of the next section, also their data do not show evidence of gravitational redshift.

\section{Analysis} 

\begin{figure}
\centering
\includegraphics[width=9cm]{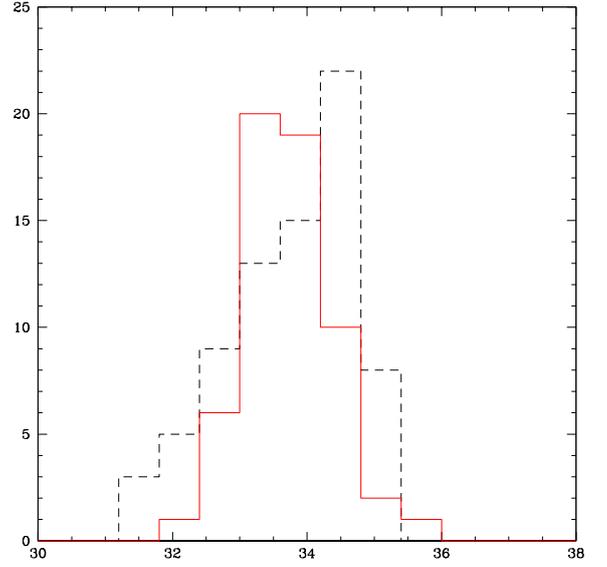}
\caption{Distribution of the measured radial velocities (RV) for the sample stars, sub-divided into 'giant' and 'dwarf' sub-samples.  Giants include all evolved stars with B$-$V$>$0.7.  No difference is present in the central value of the two distributions.  The dwarf distribution (dashed black line), shows a larger spread than the giants (904 vs.\ 680 m~s$^{-1}$) } 
\label{fig:d6469_d6456}
\end{figure}

After cleaning the sample, we performed a gross split between 'main sequence' and 'giants', defining as giants all stars along the subgiant and red-giant branches redder than $B-V$=0.7, and all the other stars as belonging to the main sequence.  The giants' radii are so large that gravitational redshift is expected to be negligible in these, and therefore their average RV should be bluer than the main sequence stars by several hundred m~s$^{-1}$.  This simple split of the sample  however does not produce the expected result, as evident from  Figure~\ref{fig:d6469_d6456}:

\begin{enumerate} 

\item Contrary to what is expected from gravitational redshift predictions, the two distributions  are centered on almost the same velocity (33.75~km~s$^{-1}$) 
\item Main sequence stars and giants have substantially different velocity dispersions:
 $\sim$~0.90 and 0.68~km~s$^{-1}$ respectively. 
 
\end{enumerate}

In proceeding with a more complete analysis, masses and radii were computed for all stars.  While the mass range is small, the radii vary considerably.  These were computed using stellar bolometric magnitude and effective temperature, calculated from the $B-V$ vs. T$_\mathrm{eff}$ relation and the bolometric magnitudes of Alonso et al.\ (1996, 1999).  A reddening of 0.041 (An et al.\ 2007; Taylor 2007) and a distance modulus of 9.61 (Pasquini et al.\ 2008) for the cluster were assumed.  For the masses, a fit between mass and luminosity was derived from evolutionary isochrones (Girardi et al.\ 2000) assuming solar metallicity and solar age.  For the evolved stars, a flat initial mass of 1.32 M$_{\odot}$ was adopted (however, small imprecisions in masses or radii do not influence our conclusions). 

\begin{figure}
\centering
\includegraphics[width=9cm]{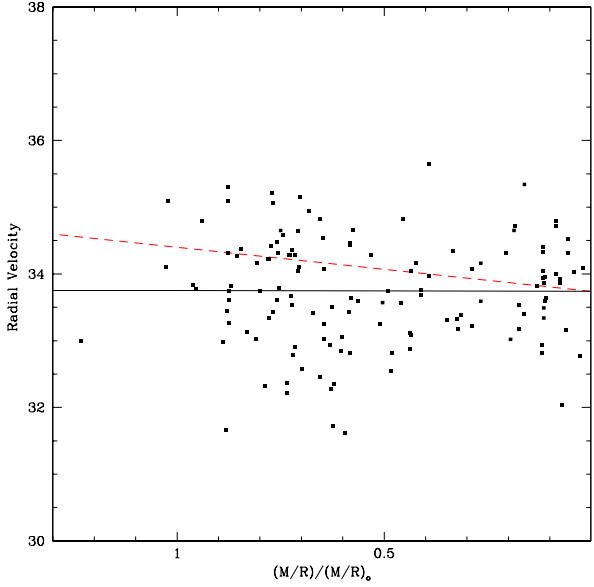}
\caption{Examining the gravitational redshift predictions: Radial velocity vs.\ the mass/radius ratio ($M/R$).  The black line is the fit to the data, while the dashed [red] line shows the linear dependence  upon $M/R$ expected from the gravitational redshift formula.  Such a dependence is not seen, at least when all the stars are considered.  Limiting the analysis to the 'main sequence' objects ($M/R > 0.5$), a trend (though not statistically significant) might be present. } 
\label{fig:GR}
\end{figure}

The result is in Figure~\ref{fig:GR}: RV as a function of the mass-to-radius ($M/R$) ratio. If gravitational redshift were the only mechanism acting, one would expect a linear relationship, which is not present: the overall slope of the observed data can actually be fitted with a coefficient consistent with zero. 

A closer analysis suggests that a trend among the dwarfs (e.g., $M/R >$0.5) might be present, fitted by a RV = 0.99($M/R$) law.  However, given the small range in $M/R$ (a factor 2 only) covered by the main-sequence stars, the uncertainty in this slope is large, and the slope is not statistically significant.  Giants, if analyzed alone, show instead a negative slope of RV vs.\ $M/R$, also with a large error.

\subsection{Maximum likelihood estimation of the kinematic cluster parameters}

Following the approach laid out by Pryor \& Meylan (1993) we performed a maximum likelihood estimation (MLE) of the kinematic cluster parameters assuming a simple model including the effect of the gravitational redshift, and separating the two stellar groups of giants and dwarfs.  We are interested in the cluster systemic velocity \vc, the velocity dispersion of giants and dwarfs, \sigc\ and \sigd, respectively, as well as the dependence of the spectroscopically measured velocity on (M/R) -- presumably due to the gravitational redshift.  We assumed that each individual measurement $i$ was drawn from a probability density
\begin{equation}
l_i(\vc,\alf,\varx) = \frac{1}{\sqrt{2\pi(\varx+\vare)}}\,
\exp\left(-\frac{(\vi-\vc-\alf\pai)^2}{2(\varx+\vare)}\right).
\end{equation}
\vi\ is the spectroscopically measured radial velocity, \sigx\ the velocity dispersion of the group of stars the object~$i$ belongs to, \sige\ the measurement error on a single star for which we assumed it is the same for all stars and amounts to 0.3\,km\,s$^{-1}$.  For brevity, we introduced $\pai\equiv(M/R)_i$. \alf\ is a constant of proportionality which should come out as 0.636\,km\,s$^{-1}$ per unit $M/R$ (in solar units) if the expected gravitational redshift was present\footnote{Here, we neglected the 3\,m\,s$^{-1}$ redshift in the gravitational potential of the Sun at Earth orbit.}.
We assumed that there is no measurement uncertainty on the \pai.  This approximation is justified since the uncertainties are dominated by the spectroscopic ones which correspond to an uncertainty in \pai\ of $\sim 0.5$, much worse than the precision expected for \pai-values from isochrone fitting.

We maximised the overall likelihood
\begin{equation}
L(\vc,\alf,\varc,\vard) = \Pi_{i\in\{\mathrm{giants}\}} l_i(\vc,\alf,\varc) \,
\Pi_{j\in\{\mathrm{dwarfs}\}} l_j(\vc,\alf,\vard), 
\end{equation}
i.e., we assumed that giants and dwarfs share the same systemic velocity, and exhibit an intrinsic line shift proportional to their mass to radius ratio.  Giants and dwarfs can potentially have different velocity dispersions -- as is expected from the previous analysis.  The ansatz for the likelihood above emphasises the role of intrinsic line shifts in the determination of the kinematic parameters of the cluster, and the need for correcting for intrinsic shifts if one aims at a precise determination of
the kinematic properties of cluster members. 

We followed the standard procedures for determining the most likely parameters and their covariance matrix, except that we used the variances instead of standard deviations as actual parameters in the minimisation process.  This had mostly practical reasons since some algebraic expressions became much more
compact this way.  For a single group of stars with velocity dispersion \sigx\ the best fitting parameters could be found analytically as 

\begin{equation}
\alf = \frac{\xmean{\vi\pai}-\xmean{\vi}\xmean{\pai}}{\xmean{\pai^2}-\xmean{\pai}^2},
\end{equation}
\begin{equation}
\vc = \xmean{\vi} - \alf\xmean{\pai},
\end{equation}
and
\begin{equation}
\varx = \xmean{(\vi-\vc-\alf\pai)^2} - \vare.
\end{equation}
\xmean{.} denotes the (ensemble) average.  We did not find an analytical solution for the case of two different velocity dispersions. For this case we solved the minimisation problem numerically by direct iteration on \varc\ and \vard\ as suggested by Pryor \&\ Meylan (1993).  With the best fitting parameters we evaluated the covariance matrix $(-\partial^2 L 
/\partial\theta_i\partial\theta_j)^{-1}$ where $\mathbf{\theta}$ is the vector
of parameters  $\mathbf{\theta}=(\vc,\alf,\varc,\vard)$. We obtained the best
fitting parameters and their $1\,\sigma$ uncertainties as

\begin{equation}
\begin{array}{llrll}
\vc    & = &  33.824 & \pm\,0.121 &\mbox{km\,s$^{-1}$}\\ 
\alf   & = & -0.134  & \pm\,0.230 &\mbox{km\,s$^{-1}$}\\
\sigc  & = &  0.566  & \pm\,0.068 &\mbox{km\,s$^{-1}$}\\
\sigd  & = &  0.849  & \pm\,0.078 &\mbox{km\,s$^{-1}$}.
\end{array}
\end{equation}

The errors of the standard deviations were calculated from the errors on the variances applying the standard error propagation formula.  From the covariance matrix we calculated the correlation coefficients among parameter pairs: $-$0.840 (\vc, \alf); $-$0.067 (\vc, \varc); 0.059 (\vc, \vard); 0.093 (\alf, \varc); $-$0.081 (\alf, \vard); and $-$0.008 (\varc, \vard).  The only significant and actually strong correlation among the parameters is the one between \vc\ and \alf.  This is understandable since changes in the systemic velocity can to some extent be traded against changes of the intrinsic line shifts. The velocity dispersions in each group of stars are largely insensitive to the systemic velocity and the intrinsic line shifts.  All in all, the MLE analysis confirms our findings that there is a clear difference ($3.6\,\sigma$) between the velocity dispersions of giants and dwarfs, and a dependence of the intrinsic line shifts on $(M/R)$ which deviates significantly ($3.3\,\sigma$) from the expected behaviour for a signal solely due to a gravitational redshift.

\section{Discussion}

Our data show two main results: that no obvious signature of gravitational redshift is seen in the data, and that the dispersion of the radial velocities is higher in un-evolved stars than in evolved ones.  We discuss these two results separately: 

\subsection{Missing gravitational redshift}
As discussed above, a radial velocity shift of up to 600~m~s$^{-1}$ is expected between main sequence stars and giants. Such a difference is not present in our data as illustrated in Figures~\ref{fig:d6469_d6456} and ~\ref{fig:GR}.  Could we have missed the gravitational redshift because of some systematic effect?

\subsubsection{Template effects} 
Given that the mask used is of spectral type K0~III, a better match between
template and target star is expected for the sub-giants and giants than for main sequence stars. 
One clear effect of using the K0~III mask with G-type turn-off stars is that the resulting CCF is shallower than for the giants and sub-giants (Figure 3 of Paper I).  This is both because of the mismatch between star and template, and because of the generally weaker lines in dwarf stars; the depth of the CCF is thus related to the luminosity.

\begin{figure}
\centering
\includegraphics[width=9cm]{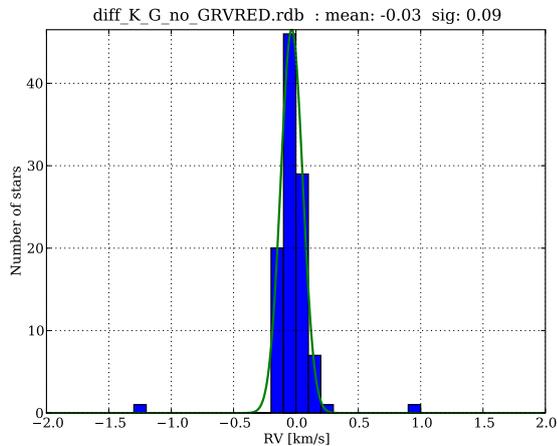}
\caption{Histogram of the differences of radial velocities obtained by using two different 
cross-correlation masks (K0~III giant and G2~V dwarf): $RV_{K}-RV_{G}$. The Gaussian fit gives a mean of $-$30~m~s$^{-1}$ and an RMS of 90~m~s$^{-1}$.} 
\label{fig:diff_histo}
\end{figure}

Could the mismatch between luminosity class and mask have caused a velocity shift in the turnoff stars? To investigate this, we computed new cross-correlation functions to all stars using a G2~V mask instead.  The mean RV difference was found to be a well defined value (30~m~s$^{-1}$) due to template zero-point mismatch. The RV difference distribution has an RMS of 120~m~s$^{-1}$.  A detailed inspection of spectra showing differences larger than 100~m~s$^{-1}$ reveals that low S/N along with a not optimum cosmic ray cleaning are responsible.  In addition, most of the discrepant spectra were recorded during the last season, when a simultaneous Th-Ar lamp was used, contaminating faint spectra through its emission lines.  If only data from the first two seasons are considered, the RMS drops to 90~m~s$^{-1}$, as in Fig.~\ref{fig:diff_histo}. 

While these tests show some limits in the precision, the use of a different mask also shows that there is no offset introduced by the choice of the mask differentially affecting various groups of stars.  However, an extra error of about 100~m~s$^{-1}$ due to poor cosmic-ray cleaning and ThAr contamination has to be added to the formal error of the cross-correlation fit. 

We also tested different mask characteristics, such as dependence of the measured RV upon the wavelength range, finding that values retrieved from the 500-700 nm mask are within a few tens of m~s$^{-1}$ from those found for the 'full' 400-1000 nm one, for different types of stars.  We therefore conclude that the mask and wavelength range used do not induce systematic effects which could cancel the gravitational redshift signal.

\subsubsection{Spectrograph drifts} 
Variations in temperature and atmospheric pressure during the observing night can cause spectral drifts of a few hundred m~s$^{-1}$ (e.g., Platais et al.\ 2007).  This effect is seen in our data collected in the third season, where the simultaneous calibration fiber was used.  Unfortunately, the first two datasets did not make use of the simultaneous calibration fiber and the spectrograph drifts could not be removed.

In order to estimate the amplitude of the drifts caused by changes in atmospheric conditions we used 52 observations of $\tau$~Cet (HD~10700) taken from the SACY survey (Torres et al.\ 2006) and from our own observations.  These spectra were reduced and analyzed in a similar way to the present M67 stars.  They cover a time span of 3000 days from Oct.\ 1999 to Feb.\ 2008.  The mean RV is $-$16.37~km~s$^{-1}$ with an RMS of 0.3~km~s$^{-1}$.  The best fit, excluding 6 extreme values, gives a similar mean, and a sigma of 0.18~km~s$^{-1}$.  Thus we believe that the drift error is about 200~m~s$^{-1}$.

\begin{figure}
\centering
\includegraphics[width=8.0cm, height=8.0cm]{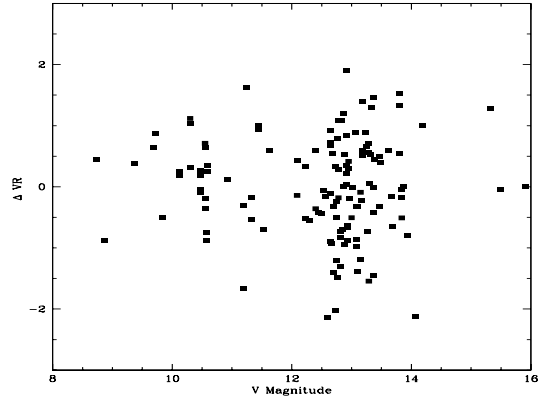}
\caption{Residuals of measured radial velocity VR as function of stellar apparent magnitude.  The residuals show a typical increase with magnitude as expected from increasing noise for fainter stars.  The effect  might also be due to a dynamically hotter population of dwarfs, or to some observational bias. } 
\label{fig:pippo}
\end{figure}

Simulations of the effect of photon noise in the RV computation show that already with a S/N of 10--30, radial velocity errors are as low as a few tens of m~s$^{-1}$ (Paper I; Bouchy et al.\ 2001).  However those simulations only consider white (photon) noise, and (as discussed above) there are additional sources of errors associated with low S/N.  In Figure~\ref{fig:pippo} we plot the scatter diagram as a function of stellar magnitude and indeed the behavior is that expected, if the noise was not negligible.

We investigate if one can reproduce the missing gravitational trend and the scatter seen in Figure~\ref{fig:pippo} accounting for the different sources of spread in the VR.  First, the intrisic velocity dispersion of the cluster of 0.8~km~s$^{-1}$ was considered.  A Gaussian drift error drawn from a distribution with an RMS of 200~m~s$^{-1}$ and a photon noise of 150~m~s$^{-1}$ for a V=14 star were added. This noise was scaled as the square root of the flux for a given magnitude (i.e., $1/SN$, assuming that photon noise is the limiting factor for faint targets). 

We then computed the difference of the center of the RV distribution of main sequence and giant stars: this is always below 100~m~s$^{-1}$.  Moreover, the RMS of both distributions is clearly dominated by the cluster velocity dispersion.  An additional gravitational shift of 600~m~s$^{-1}$ added to the distribution of sub-giant and giant stars can be retrieved in all cases in spite of their limited number.

\begin{figure}
\centering
\includegraphics[width=8.0cm,height=8.0cm]{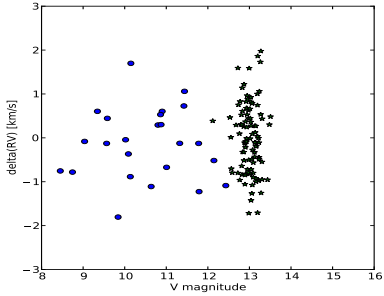}
\caption{Synthetic reproduction of the observed residual RV vs.\ magnitude diagram of Figure~\ref{fig:pippo}.  As input for the points, our various sources of noise were used, together with the velocity dispersion inside the cluster. } 
\label{fig:scatter_sim}
\end{figure}

We cannot so far identify any instrumental or observational effect which could cancel out the expected 
gravitational redshift signal, and we conclude that our observations show the absence of evidence for a 
gravitational redshift trend. 

Nordstr\"{o}m et al.\ (1997) briefly discuss a similar case, because they found a difference of 0.4~km~s$^{-1}$ (dwarfs redder than giants) between the average RV of giants and main sequence stars in the open cluster NGC~3680, almost exactly what is predicted from gravitational redshift.  NGC~3680 is an almost completely evaporated cluster, with only a few bona-fide members left.  The paucity of the sample precludes certain conclusions, but could suggest that our results about M67 may not be generalized to other clusters.  On the other hand, the fact that no gravitational redshift is found by analyzing the published data by Mathieu et al.\ (1986) shows that this result is not limited to our observations only, but it is a characteristic of this cluster. 

\subsubsection{ Cluster dynamics} 

\begin{figure}
\centering
\includegraphics[width=9cm]{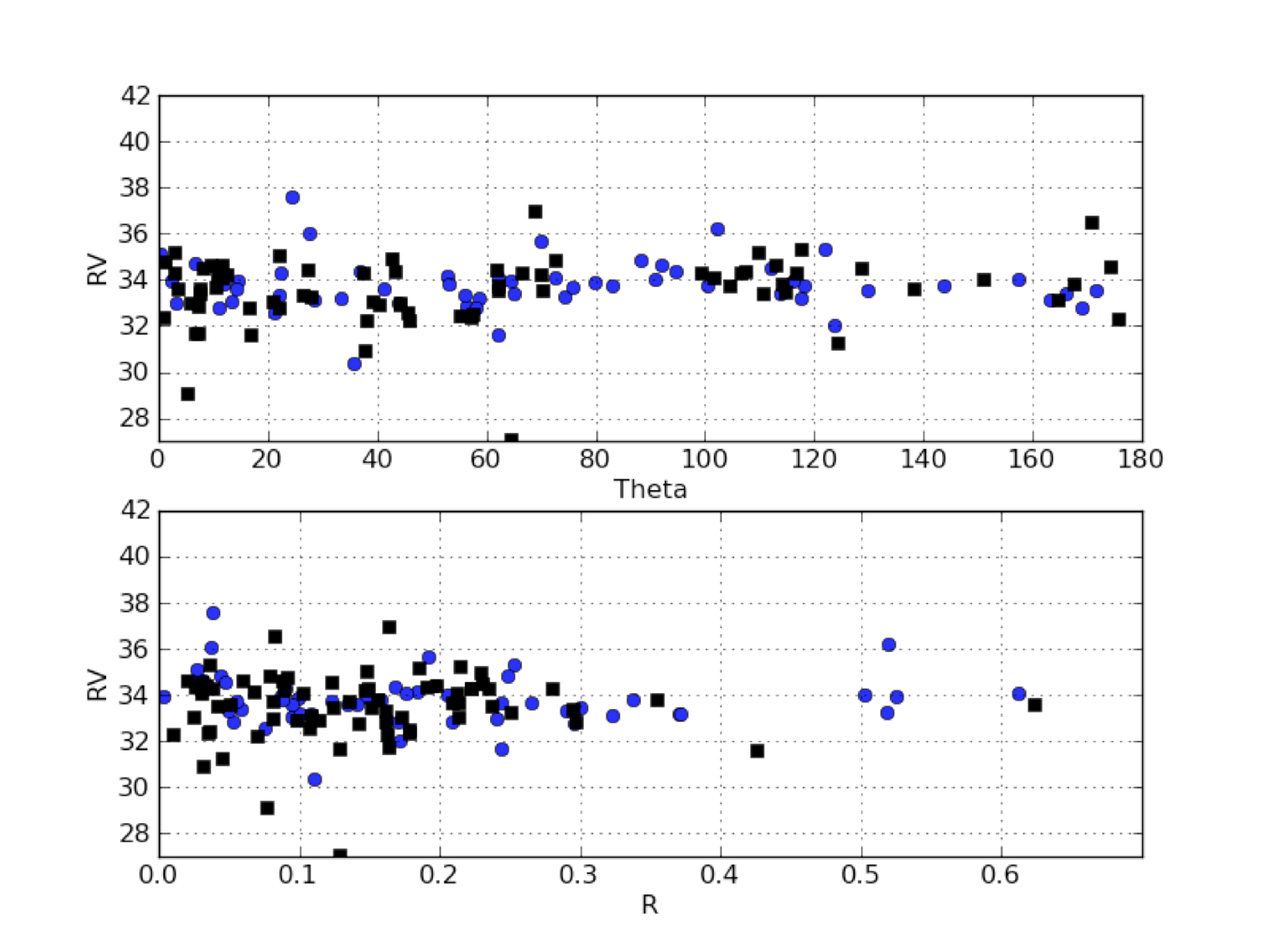}
\caption{Angular and radial distribution of measured radial velocities (black squares denote giants; [blue] 
dots are main sequence stars).  No evidence of complex dynamics or dynamical heating at larger distances
is observed.} 
\label{distrib}
\end{figure}

Even if open clusters are not known to rotate or to have complex dynamical topologies, we shall analyze whether the missing gravitational signature could be due to a combination of complex cluster dynamics and bias selection, namely that the observed giants and dwarfs are situated in different spatial regions of
the cluster, perhaps affected by different global velocities.  In this case the stars would not share the same systemic velocity and this bias could conspire to cancel the gravitational redshift signature.  For M67, there is no report of complex dynamics, with the exception of a claim by Loktin et al.\ (2005) of a possible expansion of the cluster core.  No evidence of rotation is found in the analysis of the Mathieu et al.\ (1986) data by the same authors.  In order to investigate effects related to cluster dynamics we have analyzed the velocity distribution of the observed stars as a function of R (distance from the cluster center) and $\theta$ (angular position) for giants and dwarfs, finding no evidence of RV variations with these parameters (Figure \ref{distrib}).  We note in passing that we do not find evidence for a larger dispersion at large radii.  We therefore conclude that complex cluster dynamics coupled to a selection bias cannot explain the non-detection of gravitational redshift.

\subsubsection{Convective shifts}
 
Dynamic effects in stellar atmospheres may produce net shifts in photospheric absorption lines, and we cannot exclude that lines in main-sequence stars could be more blueshifted than in giants, thus (partially) canceling the expected gravitational redshift signature.

Such a differential shift need not be present in all spectral lines, but only (in a statistical sense) in the deeper ones.  In fact, for a given metallicity, typical lines (metallic neutral ions) are generally deeper in cool giants than in dwarfs.  By using the same mask for all stars, we use lines with different equivalent widths in the two classes of stars, and relatively deeper in the giant spectra. 

\begin{figure}
\centering
\includegraphics[width=9cm]{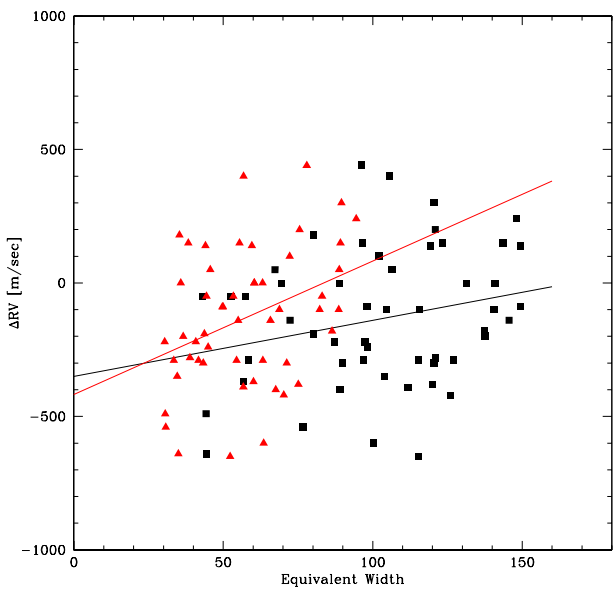}
\caption{Difference between measured radial velocities of selected lines in the median 'giant' (black squares) and 'dwarf' spectra ([red] triangles).  A trend of $\Delta$RV vs.\ equivalent width [m{\AA}] is present: stronger lines show smaller (negative) differences between dwarfs and giants.  For strong lines, the expected gravitational redshift probably is canceled by atmospheric dynamical effects. }
\label{fig:vrdiffgd}
\end{figure}

Since single spectra of each star have a rather low S/N ratio, we have constructed two 'average' low-noise spectra: one from 64 selected dwarfs (called 'dwarf') and one for 20 selected giants ('giant').  The spectra were first normalized in flux at 600 nm, and then averaged by calculating the medians.  Given the low velocity dispersion of the cluster stars, these co-added spectra are not appreciably broadened.  The S/N ratio of these spectra is about 200 in the region of interest for the 'dwarf' spectrum and about 140 for the 'giant' one.  The co-added spectra still show -- within the uncertainties -- the same radial velocity, for 'giant' and 'dwarf' alike.  We selected a number of lines used for chemical analysis of solar metallicity stars (Pasquini et al.\ 2004), inspecting them visually in the co-added spectra.  Out of the original list, 52 lines passed this quality check, and were fitted with a Gaussian function, using the ARES package (Sousa et al. 2008).  We first tested with the IRAF task Splot the consistency between the automatic program and the manual fitting.  The central wavelengths in the 'dwarf' and 'giant' spectra were recorded, and the DV = V$_\mathrm{giant}$ -- V$_\mathrm{dwarf}$ is plotted in Figure \ref{fig:vrdiffgd} as function of line equivalent widths.  For 'giant' lines weaker than about 4~pm (40 m{\AA}), the measures were not very reliable, and such weak lines are not used.  On the other hand, lines with equivalent widths above 15~pm (150 m{\AA}) were discarded to minimize  possible contamination by blends.  In both plots of Figure \ref{fig:vrdiffgd} a trend is present: the stronger the line, the redder the shift in the giant spectrum.  The observed relative shifts are up to $\pm$500 m~s$^{-1}$, comparable to the expected gravitational redshift.  We looked for, but did not find any dependence with other quantities (e.g., wavelength).

Formal fits to the trends give: DV=2.1*EW$_\mathrm{dwarfs}-$350~m$s^{-1}$ and DV=5.0*EW$_\mathrm{giants}-$418~ms$^{-1}$.  The trends are only marginally significant (correlation coefficients 0.25 and 0.35).  The formal error in measurements of single lines is about 50 ms$^{-1}$, and we believe that the observed spread is real. The lines used are mostly from neutral elements (dominated by Fe) with different characteristics, but also a few \ion{Fe}{II}.  With higher S/N spectra for single stars, it would be possible to further investigate trends between different chemical elements and line parameters, but we believe that for our current data this would be an over-interpretation.

Wavelength shifts induced by convective motions have in particular been studied for lines in the Sun and in Procyon, where it has been possible both to measure and to model intrinsic lineshapes under very high spectral resolution (Dravins 1982; Allende Prieto et al.\ 2002a); a successively increased blueshift for weaker lines is a signature of the three-dimensional photospheric hydrodynamics (e.g., Asplund et al.\ 2000).  Also, in other cool stars, Allende Prieto et al.\ (2002b) and Ram{\'{\i}}rez et al.\ (2008) observed how lines from neutral species get more blueshifted as they become fainter.

Since we are able to compare dwarfs and giants of the same cluster, some information can be added: the trend seems to differ between main sequence and evolved stars.  In the M67 evolved stars, the blueshift is less than in turnoff ones, and the expected gravitational redshift is seen only for a few of the faintest lines.  The blueshift variation with equivalent width is larger in giants than in dwarfs, as suggested by the trend in Figure \ref{fig:vrdiffgd}.  A caveat, however, is that further parameters may enter, in particular line broadening by stellar rotation which differs among dwarfs and giants: rotational broadening changes line asymmetries and may affect the velocity signal measured by cross correlation (Dravins \& Nordlund 1990).

This discussion suggests a plausible explanation for our non-detection: gravitational redshift is largely canceled because our correlation mask is dominated by rather deep lines, which are more blueshifted in main sequence stars than in giants because of photospheric convective motions. 

\subsection{Convective line shifts in 3D model atmospheres}

In order to investigate the apparent absence of gravitational redshift, synthetic spectral lines were computed from 3-dimensional and time-dependent model atmospheres.  We synthesized 15 artificial \ion{Fe}{I} lines, spanning the range of excitation potential and equivalent width of the observed ones. All lines were placed at a wavelength of 620 nm, which is the average of the observed \ion{Fe}{I} lines in the range 586--672.7~nm.  The 3D model atmospheres providing the thermal and kinematic information for the spectral synthesis were taken from the {\sf CFIST} 3D model atmosphere grid (Ludwig et al.\ 2009). For representing the 'dwarf' and 'giant', we chose for each a single 3D model atmosphere.  Table~\ref{tab:3dmodels} summarizes their basic properties.

\begin{table*}[t]
\caption{3-dimensional radiation-hydrodynamics model atmospheres.
\label{tab:3dmodels}}
\begin{center}
\begin{tabular}[t]{lllllllllllll}
\hline\noalign{\smallskip}
Model   &  \Teff & \logg   & \moh & \lx,\ly & \lz & \Nx,\,\Ny & \Nz & \nobm & $T$ & \dirms & $v_\mathrm{rms}$ & Modelcode\\
\mbox{} & [K]    & [$\mathrm{cm/s^2}$]& & [Mm]& [Mm]&         &     &       & [h] &   $\%$    & [km\,s$^{-1}$]\\
\noalign{\smallskip}
\hline\noalign{\smallskip}
dwarf     & 5\,782  & 4.44 & 0.0 & 5.6 & 2.27  &  140 & 150 & 12  & 1.2 & 14.4 & 2.24 & d3gt57g44n58\\
giant     & 4\,968  & 2.50 & 0.0 & 573 & 243   &  160 & 200 &  5  & 152 & 18.9 & 3.04 & d3t50g25mm00n01\\
\noalign{\smallskip}
\hline
\end{tabular}
\end{center}
\tablefoot{ ``Model'' is the
    model's name used in this paper, \Teff\ the effective temperature,
    \logg\ the gravitational acceleration, \moh\ the metallicity, $\lx=\ly$ the linear horizontal size of
    the square-shaped computational box, \lz\ its vertical extent,
    $\Nx=\Ny$ the number of grid points in the horizontal directions, \Nz\ the number in vertical direction,
    $T$ the duration of the simulated time series,
    \nobm\ the number of equivalent frequency points considered in the solution of the radiative     transfer equation, 
    \dirms\ the relative spatial white light intensity contrast at stellar disk center,
     $v_\mathrm{rms}$ the temporal and spatial average of the vertical velocity at Rosseland optical depth unity,
    ``Modelcode'' is an internal identifier of the model sequence.}
\end{table*}

The wavelength shifts of the synthesized lines were measured applying the same procedure as for the observed ones, i.e., by fitting a Gaussian profile.  In contrast to the observations, the artificial lines were available at very high spectral resolution ($\approx 2\times 10^6$) and signal-to-noise ratio so that their shifts with respect to the laboratory wavelength could be calculated with high fidelity.  For the 32 observed \ion{Fe}{I} lines we calculated convective line shifts by interpolating in the grid of synthetic lines according to the observed equivalent width and excitation potential of the line.  The typical statistical uncertainty to which the wavelength position of an individual line is predicted is (from previous experience) $~30\,\mbox{m\,s$^{-1}$}$.  This means that for the difference between wavelength shift of the same line between two models we expect a typical precision of $~42\,\mbox{m\,s$^{-1}$}$.  The statistical uncertainty is related to issues like how well the acoustic oscillations present in the computational box are averaged out.  The example illustrates that we further expect that the line shifts among lines of the same model are correlated, hence we do not expect that the uncertainty of the predicted shifts among giant and dwarf drops as the square-root of the number of lines used to characterise it.  Thus, we expect a modeling uncertainty of about $\pm 40\,\mbox{m\,s$^{-1}$}$ for the line shift difference between the giant and dwarf -- not a large value but also not completely insignificant.
 
\begin{figure}
\centering
\includegraphics[width=\columnwidth]{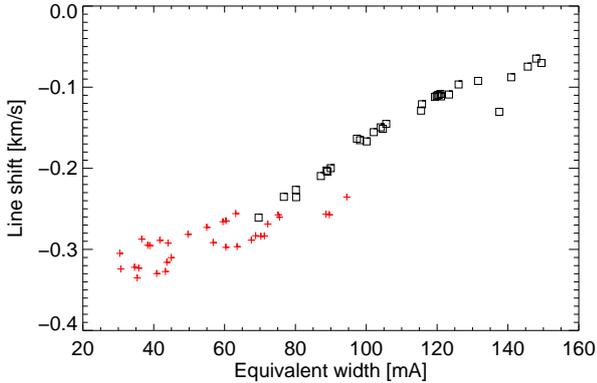}
\caption{Calculated convective wavelength shifts for the 32 studied \ion{Fe}{I} lines for the dwarf (crosses) and giant (squares) models.  Negative velocities correspond to absolute blueshifts (gravitational redshifts are left out). }
\label{fig:EW-RV}
\end{figure}

Figure~\ref{fig:EW-RV} shows the predicted convective shifts in dwarfs and giants.  Before determining the line shift, the synthetic spectral line profiles were broadened with a Gaussian profile of $\sigma$ 3.9\,km\,s$^{-1}$ which corresponds to the spectrograph resolution, combined with the macroturbulence-like velocity dispersion stemming from the motions of individual cluster stars.  In addition, it was assumed that the dwarf rotates with a projected rotational velocity of 5\,km\,s$^{-1}$ while the giant exihibits no rotation.  We find for the difference of the shifts of lines between the giant and dwarf a minimum of 62\, a maximum of 211\, and an average of $146\,\mbox{m\,s$^{-1}$}$.  We note that the correlation between the strengths of lines in the giant and dwarf is not very tight.  However, as seen in Fig.~\ref{fig:EW-RV}, we find a trend of the convective shift with equivalent width which is continuous between dwarfs and giants: on average the stronger \ion{Fe}{I} lines are less blueshifted in the giant.

\begin{figure}
\centering
\includegraphics[width=\columnwidth]{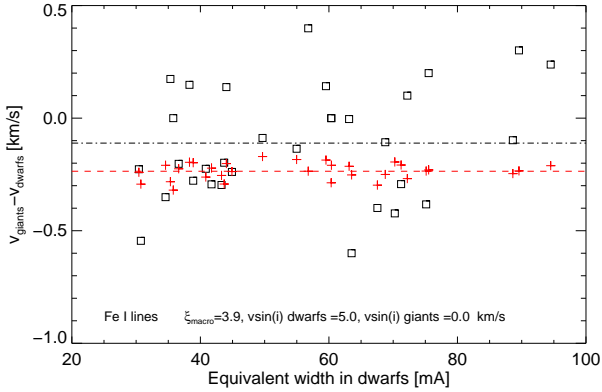}
\caption{ Comparison of observations to synthetic wavelength shifts of \ion{Fe}{I} lines, computed from 3-D hydrodynamic models for dwarfs and giants.  The computed differences between the lineshifts in giant and dwarf models is shown as crosses, and the observations as squares.  The calculated shifts include the average expected contribution of the gravitational redshift of 382~m~s$^{-1}$ }
\label{fig:FeIsample2}
\end{figure}

Figure~\ref{fig:FeIsample2} depicts the calculated differences of the shift of the same line between dwarf and giant, comparing them to observations.  For the synthetic lines, assumptions about spectrograph resolution and stellar rotation were the same as before.  An average gravitational redshift was now included, so that the giant synthetic points were blueshifted by 382\,\mbox{m\,s$^{-1}$} (applying an average M/R of 0.73 for dwarfs and 0.13 for giants, as for the stars used to compute the average spectra).  The agreement with the observations is now quite good, with a difference of only $\sim 140\,\mbox{m\,s$^{-1}$}$ between the synthetic and the observed average shifts.

However, the observed spectra show effects not predicted by the modeling: dependence of the shift on the equivalent width, and a considerable spread in the measurements.  Nevertheless, we conclude that convective line shifts are the most likely explanation as to why the expected gravitational redshift signature was not observed.

Among possible causes of remaining discrepancies between models and observations, we note that the chosen 3D models may not be the optimal representatives of the 'dwarf' and 'giant' spectra.  Their choice was dictated by availability.  For example, the iron abundance determined from the dwarf model is approximately solar but for the giant it turns out to be 0.5\,dex greater.  This may indicate that the giant model is hotter than the typical observed giant.  For the dwarfs, solar temperature is appropriate for our fainter stars, while higher temperatures are expected for the stars closer to the turnoff.  Indeed a similar Fe abundance is obtained by adopting 6000 and 4600 K for dwarfs and giants respectively (using 1D models).  However, we do not expect the line shifts to sensitively depend on stellar parameters (as long as the line strength is matched) so we argue that our approach should give a reasonably accurate estimate of the expected line shift differences.

Another possibility is the presence of noise and biases in our combined spectra; in order to test that we repeated the analysis using the Kitt Peak solar flux spectrum (Kurucz et al.\ 1984) and the Arcturus spectrum (Bagnulo et al.\ 2003).  Although here we can only test the slope and the spread around the relationship, this gives qualitatively similar results to M67, although the slope is shallower ($\sim$ 2~m~s$^{-1}$/m{\AA} instead of 5) and the scatter is smaller (100~m~s$^{-1}$ instead of 240~m~s$^{-1}$). 

Being the first attempt of such a test, we believe it is encouraging.  In presence of higher S/N data it will be possible to test other effects, such as the dependence of the line shifts on other parameters (element, excitation potential, ionization, wavelength).  One could also apply the same technique to clusters of different metallicity and chemical composition, and to work out comparison grids for theoretical models.

\subsection{Dynamically hot dwarfs}

The observation that dwarfs are dynamically hotter than giants might seem counter-intuitive, because giants are known to be intrinsic RV variables, with variations of up to a few hundreds m~s$^{-1}$ (e.g., Setiawan et al.\ 2004), and their intrinsic RV variability will add to their RV dispersion.  In addition, main sequence stars have lower masses, and in case of mass segregation the giants should be (as they are in M67) mostly concentrated in the middle of the cluster, near the center of the potential well.  On the other hand, in an equipartition regime, giants should be dynamically cooler, because they are more massive. 

The proper-motion dispersion of the bright M67 stars as measured by Girard et al.\ (1989) is almost 0.2 mas/yr.  At a distance of 836 pc (Pasquini et al.\ 2008) it corresponds to a radial velocity dispersion of 0.8~km~s$^{-1}$, comparing very well with the dispersion of RV=0.83~km~s$^{-1}$ measured from all stars and also with the 0.68~km~s$^{-1}$ dispersion of the giants only.  The comparison of the radial velocity distributions for both populations in Figure~\ref{fig:d6469_d6456} suggests that the dwarfs are dynamically hotter than the giants in M67, i.e., they have a larger velocity dispersion.

In order to check whether this might originate by some observational bias, we performed the following test: we again simulated the radial velocity distribution for the dwarfs using the same error parameters and a population of 100 stars as in Figure~\ref{fig:scatter_sim}.  The RMS of this artificial distribution was then computed, and the procedure repeated 5000 times.  The result is a distribution of RMSes, centered at the velocity dispersion of the cluster (800~m~s$^{-1}$) with values between 0.6 and 1.0~km~s$^{-1}$.  If in the simulations we now turn off the spectrometer drift error (Sect.\ 6.1.2), while maintaining all other parameters, we end up with similar figures.  This already indicates that the observed velocity dispersion does not depend on measurement errors, but rather on the number of objects. 

\begin{figure}
\centering
\includegraphics[width=9cm]{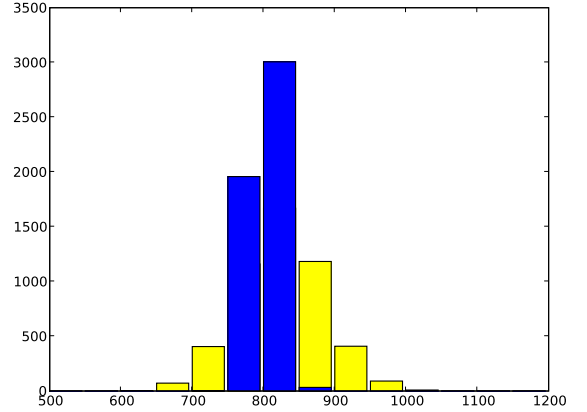}
\caption{Distribution of velocity dispersions for 5000 realizations of simulated populations of 1000 (dark) and 100 stars (light), using actual parameters for observational errors. }
\label{fig:rms_rms}
\end{figure}

As a further test, we set the drift error to its initial value and generated the RMS distributions for two 
populations: one with N=100 members (corresponding to our real case) and another with 1000 members.  The result is in Figure~\ref{fig:rms_rms}.  With 1000 members, the velocity dispersion is well defined with a center value of 0.8~km~s$^{-1}$ and 0.73 to 0.88~km~s$^{-1}$ peak-to-peak.  Therefore, the observation that the M67 dwarfs have a larger velocity dispersion than the giants seems not caused by our measurement errors but possibly is an effect of low number statistics.

We note that Nordstr\"{o}m et al.\ (1997) found a similar result for NGC~3680, with the dwarfs showing a dispersion of 0.67~km~s$^{-1}$ compared to 0.36 of the giants. Also, analysis of the Mathieu et al.\ (1986) data (which, however, contain a limited number of main sequence stars) is in perfect agreement with our findings, giving $\sigma(RV)$= 0.89 and 0.66 km~s$^{-1}$ for dwarfs and giants respectively.  

Since we cannot exclude that the effect is real, it is worth mentioning that a similar result was found by Zhao et al.\ (1996) and Loktin et al.\ (2005) from analyzing proper motion data.  Zhao et al.\ (1996), analyzing a sample of about 200 stars, found that the dispersion of proper motions of M67 bright stars (V$<13.5$ ) is significantly smaller than that of fainter objects (13.5 $<$ V $<$ 14.5), by a factor $\sim$1.3.  They also found that the brighter (more massive ) stars are more concentrated towards the 
center of the cluster.  Also Loktin et al.\ (2005) found a similar result, with a much larger sample. 
The larger sample allows them to subdivide the color-magnitude diagram into some sub-sample with giants, and another with turnoff and upper main sequence stars.  These two subsamples closely correspond to our division between giants and dwarfs.  For dwarfs they find dispersion velocities on average almost 1.3 times higher than for giants (their Table 3).
  
We also note that the tangential and radial velocity dispersions derived by Loktin et al.\ (2005) are substantially larger than our radial-velocity dispersions: 0.84 and 0.77~km~s$^{-1}$ for the giants; 1.01 and 1.04~km~s$^{-1}$ for dwarfs ('radial' in their work refers to polar coordinates in the sky plane).  Loktin et al.\ (2005) adopt a distance to the cluster of 904 parsecs.  Using our preferred distance of 831 pc, the radial and tangential dispersions become (0.77, 0.71) and (0.93, 0.96) km~s$^{-1}$, respectively for giants and dwarfs, fully compatible with the observed radial velocity dispersions of 0.68 (giants) and 0.9~km~s$^{-1}$ (dwarfs).  This result indirectly confirms the shorter distance scale for this cluster that emerges from recent works (cf.\ Pasquini et al.\ 2008).  
  
It is remarkable that in both our data and those by Mathieu et al.\ (1986), the ratio between the radial velocity dispersions of main sequence and evolved stars is 1.3, the same as found by Zhao et al.\ (1996) and Lotkin et al.\ (2005) for the dispersion between fainter and brighter stars.  Zhao et al.\ (1996) interpret these results claiming that the dynamical energy of the more massive stars has been passed in the time evolution of the cluster to the less massive objects, which are now escaping the cluster.  
 
While a full dynamical analysis of the cluster is beyond the scope of this work, the possibility of joining precise radial velocity and proper motion data seems attractive for detailed future investigations.

M67 is by now the best studied open cluster when combining both, the number of stars observed and the precision of the measurements. Substancially 
larger samples of radial velocity measurements have been gathered in globular clusters 
(see e.g. Sommariva et al. 2009), but they  are dynamically much more complex objects and with  
a larger   velocity dispersion.  
 
\section{Conclusions} 

A search was carried out to find the expected differences in gravitational redshift between the spectra of dwarf and giant stars in the M67 open cluster, whose stars should share the same systemic velocity, and where the average radial motion for different types of stars thus should produce the same Doppler shift.  However, despite adequate measuring precision, the expected signature was not seen.  Investigations of the reasons for absence of signal -- in particular following high-resolution spectral synthesis from 3-dimensional hydrodynamic model atmospheres -- indicate that likely reasons are different amounts of convective blushifts in different types of stars.  An enhanced convective blueshift in, particularly, those stronger spectral lines in dwarf stars that carry a larger weight in the cross-correlation techniques used for radial-velocity measurement, appear adequate to cancel the expected gravitational redshift.

This study thus also illustrates the limitations of cross-correlation techniques for high-accuracy  lineshift measurements.  To understand the origin of wavelength shifts that are only a small fraction of any spectral-line width, a cross-correlation measure over an extended spectral range may not be adequate, but rather one needs to examine (groups of) individually selected line profiles measured with sufficient spectral fidelity to reveal also their intrinsic asymmetries and amount of rotational broadening.  These can then be compared to synthetic spectral lines from 3-D atmospheres complete with convective blueshifts and  gravitational redshifts.  Not many current spectrometers have adequate performance for such tasks, but work is in progress towards realizing such high-fidelity spectroscopy with an accurate wavelength calibration at the largest telescopes (Dravins 2010; Pepe et al.\ 2010; Pasquini et al.\ 2010). With these instruments it will be possible to extend the present study to less explored stars, for instance 
in the metal-poor and  in the metal-rich regimes.

\begin{acknowledgements}
The 3D model of the giant star was computed at the CINECA supercomputing centre, which granted time through the INAF-CINECA agreement 2006, 2007.  
LP thanks Paul Bristow for a critical reading of the manuscript.  
DD acknowledges support by the Swedish Research Council and The Royal Physiographic Society in Lund.  
The suggestions by  the referee, Dr.  D. Pourbaix, helped to improve the manuscript. 
\end{acknowledgements}


{}

\end{document}